# A Critical Examination of the Ethics of AI-Mediated Peer Review

Laurie A. Schintler*, George Mason University
Connie L. McNeely, George Mason University
James Witte, George Mason University

*Corresponding author: lschintl@gmu.edu

*Abstract*
Recent advancements in artificial intelligence (AI) systems, including large language models like ChatGPT, offer promise and peril for scholarly peer review. On the one hand, AI can enhance efficiency by addressing issues like long publication delays. On the other hand, it brings ethical and social concerns that could compromise the integrity of the peer review process and outcomes. However, human peer review systems are also fraught with related problems, such as biases, abuses, and a lack of transparency, which already diminish credibility. While there is increasing attention to the use of AI in peer review, discussions revolve mainly around plagiarism and authorship in academic journal publishing, ignoring the broader epistemic, social, cultural, and societal epistemic in which peer review is positioned. The legitimacy of AI-driven peer review hinges on the alignment with the scientific ethos, encompassing moral and epistemic norms that define appropriate conduct in the scholarly community. In this regard, there is a "norm-counternorm continuum," where the acceptability of AI in peer review is shaped by institutional logics, ethical practices, and internal regulatory mechanisms. The discussion here emphasizes the need to critically assess the legitimacy of AI-driven peer review, addressing the benefits and downsides relative to the broader epistemic, social, ethical, and regulatory factors that sculpt its implementation and impact.

## 1. Introduction

Science is a central locus of knowledge in society and, as such, it is an inherently a socio-institutional construct. In that sense, knowledge governance and assessment in science communication is a fundamental social activity largely defined by the processes of scholarly peer review (Polanyi, 1962). Over the last half-century, scholarly peer review has undergone a digital transformation involving information technologies such as computers and the Internet (Vicente-Saez et al., 2021). Now, artificial intelligence (AI) — referring to technological systems "that can carry out tasks by displaying intelligent, human-like behavior" via computational formulae, rules, and logic (Russell & Norvig, 2021) — is being integrated into related activities to augment and automate various decisions, from selecting reviewers to eliminating studies judged to be low-quality or fraudulent (Heaven, 2018; Jana, 2019; Checco et al., 2021). Recent breakthroughs in natural language processors (NLPs), large language models (LLMs), and other generative AI technologies (e.g., ChatGPT[1]) could disrupt the peer review system further, bringing with it not only new prospects but also unprecedented concerns and challenges (van Dis et al., 2023). In fact, the stakes are so high with such technologies across the board that leading AI luminaries recently called for a moratorium on research and development involving related applications.[2]

---

[1]Chat Generative Pre-Trained Transformer

Against this backdrop, the research presented here is framed relative to the increasing attention to both the promises and perils of AI-driven peer review.  For example, AI and algorithmic decision making arguably can promote efficiency; they can also help alleviate some of the problems that confront peer review today, such as long decision and publication delays (Björk & Solomon, 2013).  Along these lines, the primary focus of many corresponding discussions and studies has been on technical considerations such as usability, accuracy, precision, and efficiency (Checco et al., 2021).  Still, important concerns have been raised at various points regarding the broader effects and implications of using AI and algorithmic decision making in scholarly activities (Araujo et al., 2020).  Notably, as never before, the public debut of ChatGPT in November 2022 brought attention to and ignited an intense and growing dialogue on the ethics of AI use in peer review.  Even so, the conversation has been confined mainly to matters related to plagiarism and authorship in academic journal publishing (e.g., Thorp, 2023; Stokel-Walker, 2023; van Dis et al., 2023; Hosseini & Horbach, 2023), while the broader epistemic, social, cultural, and societal context in which AI for peer review is positioned has been largely neglected.

AI comes with various ethical and social downsides and risks, potentially undermining the integrity of peer review and its outcomes. However, human peer review systems face many similar issues, fraught with biases, abuses, ethical lapses, and opacity (Resnick & Elmore, 2016), which frankly were already threatening the credibility of peer review in practice.  Such issues lead to questions about the relative legitimacy of AI for peer review, especially in contrast to human application. Does it solve a problem? Is it the "right" thing to do? Should its use be regulated?  What is at stake? Is AI transforming the very institution of science? Legitimacy generally refers to the congruence between the activities and behavior of an actor, such as an organization, and its institutional norms and values (cf. Dowling & Pfeffer, 1975). Accordingly, the legitimacy of AI and algorithmically-driven peer review encompasses the alignment of related practices and outputs with the "scientific ethos," comprising moral and epistemic norms, principles, and values that define "appropriate" and "good" behavior in the scientific and scholarly community.  Such issues constitute a kind of "norm-counternorm continuum" on which the legitimacy of AI use in peer review processes correspond to institutional logics and ethical practice.

The discussion here attends to such matters, critically reflecting on the legitimacy of AI-driven peer review processes, considering not just questions of technical utility but, importantly, those related to broader epistemic, ethical, and regulatory contextualizing and determinant dynamics. In doing so, we address how and to what extent AI peer review and application are consistent with the norms and values embedded in the scientific ethos, engaging conceptual and theoretical mapping as an interpretive frame (Friedman et al., 2013).  At its core, this research invokes a sociological institutionalist approach to frame and delineate critical concepts and relations, providing a structured and systemic analytical perspective and inquiry into the legitimacy of AI in peer review. Based on the analysis, we develop recommendations for future research, as well as for the evaluation and regulation of AI for legitimate and trustworthy use in peer review and other scholarly endeavors.

---

[2] https://www.wsj.com/articles/elon-musk-other-ai-bigwigs-call-for-pause-in-technologys-development-56327f

## 2. Theoretical and Conceptual Background

Legitimacy, as a social construct and delineating feature, indicates why something is valued and is a core aspect of institutions and institutionalization processes (Deephouse et al., 2017). To that end, legitimacy generally refers to the extent to which attitudes and actions are congruent with some socially acceptable norms corresponding to beliefs about appropriate behavior (Weber, 1978; Bicchieri, 2005). For purposes of this inquiry, legitimacy is addressed particularly as an organizational feature, considered in terms of the organizational engagement of AI for peer review. Accordingly, legitimacy references the manner and process in which an organization receives justification such that, broadly conceived, it captures the congruence between an organization's behavior and outcomes with accepted social principles (Parsons, 1956). In this sense, legitimacy is understood as grounded in and dependent on "cultural rule systems" in which norms and values shape behavior and perceptions about that behavior (Scott, 1995).

Four dimensions or types of legitimacy — pragmatic, moral, regulatory, and cognitive — reflect different behavioral dynamics. One important distinction is between *pragmatic legitimacy* and *moral legitimacy*. The former captures judgments about whether an organization benefits actors (it solves a problem), and the latter, whether its actions, behaviors, and practices are the "right thing to do" (they are ethically just) (Suchman, 1995). Thus, they are the essence of instrumental rationality and intrinsic value, respectively — i.e., the efficient means to an end instead of the end itself being "right" (Weber, 1978). Arguably, moral legitimacy may come at the cost of pragmatic legitimacy to the extent that "conformity to institutionalized rules often conflicts sharply with efficiency criteria and, conversely, to coordinate and control activity to promote efficiency undermines an organization's ceremonial conformity and sacrifices its support and legitimacy" (Meyer & Rowan, 1977; Waldo & Miller, 1948).

Legitimacy also reflects an integral authoritative dimension (Weber, 1978), embodying regulatory elements and dynamics. More specifically, *regulatory legitimacy* refers to the conformity of organizational action to "acceptable" standards and mandates and the "appropriateness" of such strategies in the first place (Deephouse & Carter, 2005). Regulation — comprising "hard" and "soft" laws, which include the use of legal rules backed by formal sanctions versus informal strategies such as "recommendations, guidelines, codes of conduct, non-binding resolutions, and standards" (Black & Kingsford Smith, 2002) — plays a critical role in shaping and maintaining institutional processes and structures. Indeed, institutions are in and of themselves regulatory systems. That is, they are "humanly devised constraints that structure political, economic, and social interaction" vis-à-vis codified laws and rules, beliefs, and social norms encoded through collective perceptions encapsulated in the ethos of the organization (Poon et al., 2022). Accordingly, regulatory mechanisms are not just exercised internally (i.e., intrinsic to an organization); they also are imposed externally on an organization, typically through legally-bound instruments and strategies (e.g., legislation).

Finally, *cognitive legitimacy* captures more tacit elements, such as how stakeholders make "judgments about an organization passively and not based on active evaluation" (Deephouse et al., 2017). Thus, from a cognitive perspective, institutions are legitimate when "they are understandable" rather than just desirable or practical (Pollack et al., 2012), although such ends are not mutually exclusive. For example, an organization may espouse values of openness and transparency for moral and pragmatic reasons. Thus, cognitive aspects of legitimacy reflect "the taken-for-grantedness" that an organization's activities and outcomes "achieve as they become institutionalized over time" (Alexiou & Wiggins, 2019).

In sum, the legitimacy of an organization and its practices reflects four interrelated dimensions: pragmatic, moral, regulatory, and cognitive (Deephouse et al., 2017). All are necessary and interact to ensure that institutional structures and practices are meaningful, predictable, and trustworthy (Suchman, 1995), both internally and externally.

## 2.1. The Ethos of Science and Legitimacy

Science is comprised of "individual scientists working in social groups in social institutions, exercising social values and activities" (Erduran & Dagher, 2014; Drori et al., 2003). As an institution itself, science involves social interactions, structures, and relations; it is, therefore, held to "certain values, expectations, and norms" (Kim & Kim, 2018). As such, the legitimacy of scientific and scholarly endeavors — and ultimately "trust within science and public trust in science" — depends on adherence to those social norms (Barber, 1987). From this, we can understand that the norms binding on scientists are encapsulated in the *ethos of science*, a code of conduct and collective expectations and understandings, or a social contract for scientists, indicating "appropriate" scientific and scholarly conduct. As initially delineated by sociologist Robert Merton (1942), four moral "normative pillars" have been posited as defining the ethos of science:

- Universalism (scientific claims must be objective and not based on impersonal criteria),
- Communality (science is common property derived through open, transparent and equitable communication and sharing),
- Disinterestedness (science should be driven only by epistemic interests, and scientists are accountable to their peers), and
- Organized Skepticism (scientific claims should be scrutinized by the community, subject to proof and verification).

Merton (1942) also specified a set of epistemic norms involving the integrity of the research itself. From this perspective, scientific claims should be grounded in "adequate, valid, and reliable evidence," and scientific methods and research outcomes should be logically consistent and transparent (Anderson et al., 2010). Such norms constitute the "normative habitus (or structuring of propensities and tendencies) that sets in place and gives shape to the basic values, dispositions, and character traits" that determine "good science" in the eyes of the scientific community (Leslie, 2021). Hence, there is an ethical, social, and epistemic responsibility to ensure, at least informally, that the norms are essentially compulsory — in the sense of regulatory legitimacy — given that they are "procedurally efficient" and believed to be "right and good" — in the sense of cognitive legitimacy — which more fundamentally is the distinction between pragmatic and moral legitimacy, capturing instrumental and intrinsic values respectively. Moreover, these aspects of legitimacy are central determinants in the institutionalization of science.

However, having said that, various societal trends and pressures — e.g., globalization, politicization, commercialization, hyper-competitiveness, and the digitalization of the scientific enterprise — have led to continual normative revisions and even to the specification of "counternorms," including solitariness, particularism, interestedness, and organized dogma, which provide direct counter-points to the normative pillars of the Mertonian scientific ethos (Mitroff, 1974). Even Merton recognized the practical difficulties in living up to such ideals (Kim & Kim, 2018) and, more to the point, the conduct of scientists can be viewed more realistically along a norm-counternorm continuum, as suggested above. Nonetheless, despite not always adhering to them, most scientists espouse and value the Mertonian norms as the ideal (Kim & Kim, 2018). Thus, we use Merton's ethos of science as an interpretative frame of reference.

Science reflects "an elaborate system for allocating rewards," primarily in the form of recognition and respect (Merton, 1957). It is principally a self-regulating, autonomous system guided by the norms encoded in the scientific ethos, where inappropriate behavior can invoke ostracism and reputational damage (Merton, 1942). In this respect, autonomy is an overarching principle, providing "a defense for the other values" and norms that characterize science as a social institution (Barnes & Dolby, 1970). Still, restrictions on autonomy may "be justified to prevent harm to people, society, or the environment" and also to foster socially beneficial research (Resnick, 2008; Bernal, 1939). In other words, external regulation may be in order when science and its practices and outputs conflict with societal values and expectations. (For example, there is an enormous university/science web of regulatory measures and bodies.) However, there is a general consensus within the scientific community that it is the process rather than the outcomes that should be subject to such regulatory approaches (Resnick, 2008).

In reference to peer review, these issues are enacted through the use of "experts in a given domain" to appraise, scrutinize, and critique "scientific work produced by others in their field or area of competence" (Lee et al., 2013; Jana, 2019). Peer review is the lynchpin around which the whole business of science pivots (Ziman, 1968), and related activities and functions are "embedded in the core of our knowledge generation systems" (Tennant & Ross-Hellauer, 2020) — in the institutional and social fabric of science. This includes not only peer-reviewed publications but also funding decisions. As such, peer review has a critical internal regulatory (and gatekeeping) function in the scientific enterprise, dictating whose and what research should be valued. In other words, as a crucial determinant and legitimizing feature by which the impact, originality, and quality of research are assessed (Lee et al., 2013; Jana, 2019), peer review provides "a system of institutionalized vigilance" (Merton, 1973).

Accordingly, peer review plays a decisive role in shaping the career trajectories of researchers and their participation in science more broadly. Moreover, as the primary means for certifying scientific knowledge, peer review is fundamental to scientific progress, determining what gets added to the scientific record and how science gets translated into practice (Ziman, 1968). Thus, peer review has profound implications for science and society, inextricably linked to trust in science inside and outside the academy (NASEM, 2017).

**2.2.    The Technological Impetus and Legitimacy**

In its first instance, technology is essentially about manipulating nature and, more specifically, about how humans change their environments or try to exceed their natural capacities (Volti, 2017). Given technology's influential role in society, its legitimacy must be carefully scrutinized. Here again, we look especially to distinctions between the pragmatic and moral legitimacy of technology. As earlier indicated, an entity is viewed as legitimate on pragmatic grounds when it is perceived to facilitate an organization's "self-defined or internalized goals or outcomes," capturing beliefs related to its "effectiveness, efficiency, or utility"; whereas it is legitimate on moral grounds when it is perceived to be consistent with the organization's or society's moral and ethical values, capturing "perceptions or beliefs related to morality, ethicality, or integrity" (Tost, 2010). The two may complement or conflict with each other such that, in regard to technology, legitimacy is tied to whether it solves a problem versus whether it should be used, i.e., is the right thing to do, where the former speaks to the instrumental behavior of the technology and the latter to values, beliefs, and attitudes regarding its use, i.e., to the intrinsic aspects of the technology.

From an instrumental position, technology is a neutral or universal object for humans to use to achieve progress (Feenberg, 2008). This approach resembles notions of technological determinism, positing that "the structure of the technology itself largely determines the uses of the technology, that is, that its functions follow from its form" (Kline, 2015; Postman, 2011). It also relies on "technological imperatives," typically expressed in needs-based terms due to functional requirements determined, in this case, by corresponding changes in practical needs, thus implying different structures and practices in keeping with technological determinist models (Schintler & McNeely, 2022). Moreover, technological determinism assumes that technology itself determines its cultural values and social structure rather than society doing so (Kline, 2015).

On the other hand, technological interactionist models reflect a more intrinsic approach, focusing on the ways that technology shapes and is shaped by societal interactions, emphasizing the "reciprocal interaction between technological and social change" (Schintler & McNeely, 2022). In other words, it emphasizes the interaction between humans and technology as a two-way process and relationship in which societal values, norms, and principles are front and center. In line with interactionist reasoning, science is recognized as a socio-technical system involving the dynamic interaction between social, cultural, and technological elements in corresponding institutional structures and processes. Indeed, throughout history, technology has significantly affected how science is vetted and disseminated within not only the scientific community, but society more generally, shaping and shaped by the scientific ethos at every turn. Underlying the processes of science gaining legitimacy "is the web of social relations among science, its practitioners, the knowledge it generated, and the society," with elements of rationality, truth, and "objective" knowledge widely "accepted and upheld by the society at large." Over time, science as a social institution and construct "has come to be recognized as part of the culture, a perspective, an ideology, and a viewpoint of society" (Krishna 2014).

The Gutenberg printing press, invented in the 15th century based on the earlier Chinese invention of moveable type, provides a prime illustration. The introduction and diffusion of this technology created seismic shifts in the "means used to acquire and establish knowledge of nature, the institutional setting within which that knowledge was validated and valorized, and the substantive content of that knowledge."[3] In other words, it contributed to "new social conventions, incentive structures, and institutional mechanisms" in science, emphasizing rapid disclosure and broader dissemination of scientific discoveries (David, 2008; Sarton, 2016), underscoring the instrumental benefits — and pragmatic legitimacy — of the technology. Moreover, it was "one of the early steps toward the democratization of knowledge,"[4] capturing more intrinsic elements of the technology in line with the democratizing principles embedded in the Mertonian ethos and, thus, the moral legitimacy of the technology.

Of course, today, scholarly communication in general and peer review more specifically have taken on an increasingly digital façade involving various more recent technological innovations and applications, including the internet, electronic mail, and social media. Indeed, digital technology has revolutionized scholarly communication (Ziman, 1996), so much so that some have referred to the digitalization of science as a second Scientific Revolution — or Science 2.0 (Bartling & Friesike, 2014). Such technologies have given rise to novel ways of "sharing and finding information online," in which "scientists can archive and establish priority for their research on preprint platforms and amplify their work on blogs and social networks," capturing a broad audience that extends to the public (Lee, 2022). This situation has contributed to "changes in the ways that scholars communicate with each other for informal conversations, for collaborating locally and over distances, for publishing and disseminating their work, and for constructing links between their work and that of others" and vetting research (Borgman & Furner, 2001). Thus, digital technologies have significantly altered science, transforming it into a globalized and highly networked enterprise — the "new invisible college" (Wagner, 2009) — speaking to "values of universalism and communalism that define the scientific enterprise" (Mielkov, 2023).

---

[3] https://blogs.ubc.ca/etec540sept10/2010/10/30/printing-press-and-its-impact-on-literacy/
[4] https://blogs.ubc.ca/etec540sept10/2010/10/30/printing-press-and-its-impact-on-literacy/

Still, at the same time, there are digital and knowledge divides, which arguably have deepened over time (Schintler & McNeely, 2012). Of particular relevance to questions of legitimacy in this regard are issues related to accessibility, referring to "the design, construction, development, and maintenance of facilities, information and communication technology, programs, and services so that all people…can fully and independently use them."[5] While these aspects of accessibility are logically and operationally interdependent and bounded, the point here is that it is a defining and determinant constituent of the scientific community and of related values and practices (McNeely & Frehill, 2023). Not every person or place has access to the resources, the technology, or the knowledge, skillsets, and motivations for using it in the first place, a problem that has worsened in the digital age (Wagner, 2009), and which is incongruent with the Mertonian ethos of science. Also problematic in this regard, and with broader implications for participation in scholarly publication and communication, is the global digital and knowledge divide in which economically disadvantaged countries and regions, and the individuals and groups involved, face significant challenges in participating in the production, consumption, and validation of science in digital fora and related platforms (McNeely & Schintler, 2022; Drori et al., 2003), undermining both the moral and pragmatic legitimacy of the scientific institution.

## 3. Legitimacy of AI for Peer Review

AI technology, encompassing techniques such as machine learning, natural language processing, computer vision, and robotics (Russell & Norvig, 2021), is woven more and more into peer review. Such tools have been instrumental in automating routine tasks, such as detecting plagiarism and image tampering. Moreover, the technology is being deployed to match manuscripts to journals and identify peer reviewers. Thus, AI is on the front lines of fighting scientific fraud, misconduct, and related ethical lapses in peer review and addressing various information asymmetries simultaneously.

While peer review has yet to become fully automated — i.e., where AI makes all decisions from desk rejects to accepting manuscripts for publication — research is underway to develop

---

[5] https://www.whitehouse.gov/briefing-room/presidential-actions/2021/06/25/executive-order-on-diversity-equity-inclusion-and-accessibility-in-the-federal-workforce

systems that could conceivably carry out such tasks. There are active efforts to design AI applications that can discern and evaluate research studies' quality, novelty, and impact (Vincent-Lamarre & Larivière, 2021). In this respect, technologies such as ChatGPT, a natural language processing tool grounded in generative AI technology that can carry out "human-like conversations" and produce realistic content, including text, video, and images, in a facile and instantaneous fashion, may be a game-changer. Accordingly, given how rapidly AI is advancing, a peer review system without a human (scientist)-in-the-loop could arrive sooner rather than later (Stockon, 2017). In other words, the idea (and actuality) of an omniscient "Robot Reviewer" is moving beyond the realm of imagination into reality, which could profoundly impact the legitimacy of peer review.

In light of such breakthroughs, AI will no doubt play an increasing role in peer review to streamline and speed up the process and to help address ethical issues (e.g., plagiarism) in the scientific enterprise, enhancing the pragmatic and regulatory legitimacy of related practices. At the same time, AI technology is fraught with myriad ethical and social issues and challenges. For instance, it is prone to errors, biases and discriminatory outcomes, safety hazards, the disintegration of social connections, and a lack of transparency and accountability (Anderson et al., 2018), raising questions about the moral and cognitive legitimacy of its use in this context. However, as previously mentioned, human peer review systems share many of the same problems. More to the point, corresponding processes and practices are arguably prejudiced, capricious, slow, inefficient, ineffective, opaque, adversarial, and incohesive (Ralph, 2016). Such issues are inconsistent with Mertonian norms and already undermine the legitimacy of peer review and, frankly, of science as a whole (Resnick & Elmore, 2016), prompting some to suggest that peer review faces a credibility crisis (Dasgupta, 2017) or, more to the point, a "legitimation crisis" or at least a "legitimation deficit." With attention to such affective issues, we critically examine how and to what extent AI in peer review affects the legitimacy of related practices and outcomes.

### 3.1. Thematic Delineations and Mapping

Attending to legitimacy in its various guises in relation to AI-mediated peer review systems, especially in contrast to human decision making as discussed above, means explicitly assessing its alignment with the scientific ethos, as summarized in Figure 1. Again, in this regard, a significant pragmatic issue raised in critiques of peer review is that it has become increasingly inefficient. Publishing in scholarly journals can entail long delays from submission to publication (or rejection), with the situation worsening over time (Björk & Solomon, 2013). This problem stems partly from the fact that millions of peer-reviewed papers are published yearly, and many more manuscripts are reviewed but rejected for publication (Nazar et al. 2022). Thus, it has become more difficult to find reviewers (Carroll, 2018), especially since qualified scientists tend to be "busy with their own research, teaching, supervision, life, grant proposals, reading, thinking, and conferences" (Pautasso & Schäfer, 2010). Science also has become an interdisciplinary endeavor, which poses additional challenges to securing reviewers, particularly those with appropriate cross-domain knowledge and expertise. In addition to these issues, humans have limited attention and processing capacities in the first place and, as such, inefficiencies in peer review are arguably tied to human shortcomings.

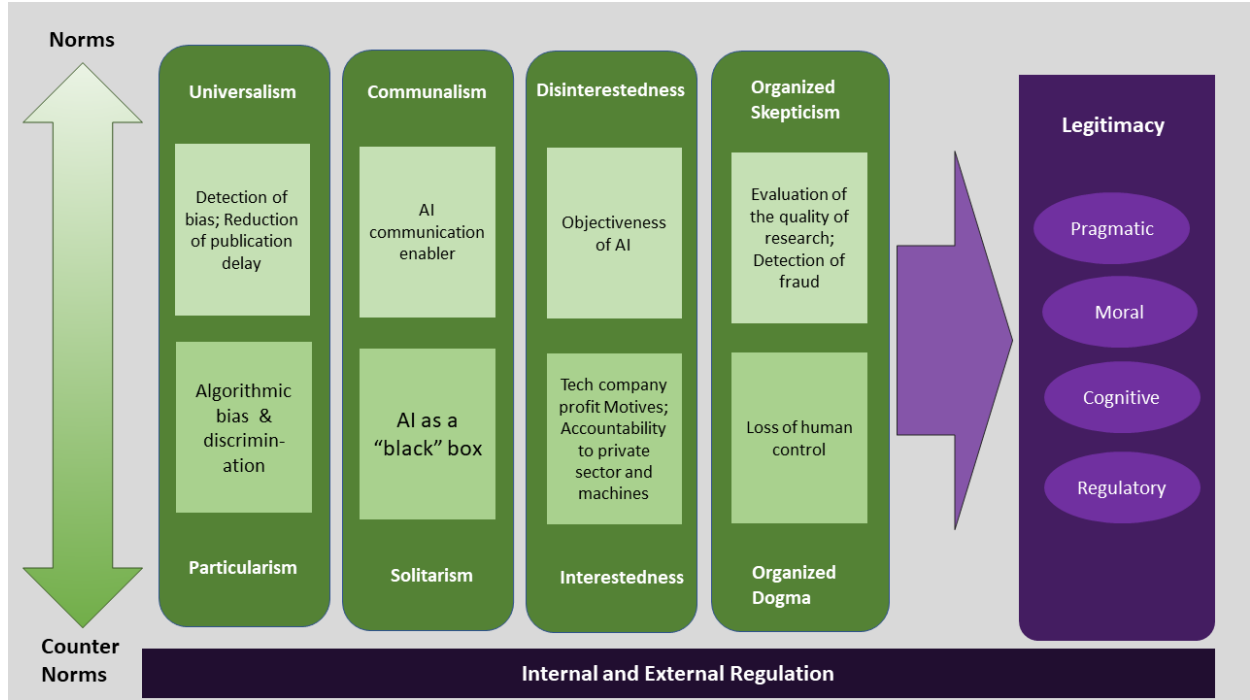

**Figure 1.** Mapping AI-Mediated Peer Review to the Ethos of Science
(Source: Author)

AI has the potential to address such concerns, ultimately making peer review more efficient and productive, bringing to the fore issues related to the pragmatic utility of the technology. Indeed, one of the instrumental benefits of AI is that it can promote efficiency, primarily through mechanisms of (fast) optimization and automation (Schintler & McNeely, 2022; Checco et al., 2020; Heaven, 2018).

*AI systems are generally capable of creating, processing, and analyzing information much faster (translated as more efficiently) — and, in some cases, more effectively and productively — than humans.*

Yet, other concerns arise alongside these advantages. Bias, for example, is another severe matter regarding peer review. The peer review system is fraught with biases, prejudices, and gatekeeping pertaining to author characteristics (e.g., institutional affiliation, discipline, gender, race, and nationality) and their research (e.g., topic, ideology, or methodology) (Lee et al., 2013). Also, cognitive biases, emotions, and similar subjective influences can (and do) play a role in peer review decisions (Street & Ward, 2019). Although the research community has started "institutionalizing practices designed to thwart human biases and increase the trustworthiness of scholarly work" (Jamieson et al., 2019), such as open peer review and single-blind review (Kovanis et al., 2017), bias remains a pervasive problem.

*Rampant bias in AI systems goes against the grain of the norm of universalism and other democratizing principles behind the scientific ethos.*

On the one hand, AI could help to mitigate biases and discriminatory outcomes inherent in peer review under the premise that the technology is more objective than humans (Araujo et al. 2020). It can even detect such problems (Checco et al., 2021). However, just like humans, AI is susceptible to bias and tends to produce outcomes and decisions that misrepresent and disfavor women, minorities, and other socially undermined individuals and groups (Anderson et al., 2018). In fact, AI algorithms have the risk of "replicating and even amplifying human biases," particularly those affecting already underrepresented groups (Lee et al., 2019). In other words, if the data used for developing such systems is statistically skewed or captures inequality or inequities present in the first place, it can learn such patterns leading to inappropriate and biased decisions. Bias also stems from the selection or, more aptly, the mis-selection of algorithms. However, even when AI is designed with "right" values in mind or using de-biased data, its application can still contribute to interpretation bias. That is, the system might be working as intended by its designer, but the user may not fully understand its utility or might try to infer different meanings that the system might not support (Galaz et al., 2021). In other words,

> *Human biases can come into play with AI such that the moral legitimacy of its use in peer review must be assessed relative to human decision-making.*

Peer review also is fraught with "filter failures" in relation to epistemic norms. That is, it is not always effective in weeding out low-quality or fraudulent research or, on the other hand, in identifying impactful research that should be published (Resnick & Elmore, 2016). Fake peer review and peer review cartels are related concerns (Ferguson et al. 2014). While AI can help to address such problems — i.e., to screen manuscripts or reviews for quality and authenticity — it also can compound matters.

> *Of particular concern is that the technology has made it possible to automate the production of manuscripts, including those that appear realistic or valid but contain inaccuracies or nonsense altogether.*

Such papers can (and do) slip through peer review systems. For example, in 2005, researchers at the Massachusetts Institute of Technology (MIT) generated a program called Scigen that could randomly generate "nonsensical computer-science papers," complete with realistic-looking graphs, figures, citations, and context-free grammar.[6] Over the years, Scigen papers and other gibberish artifacts have gone undetected under the radar of human peer reviewers, making their way to the published literature and to refereed conferences (Labbé & Labbé, 2013; Van Noorden, 2021). (Although interestingly, the original purpose of Scigen "was to expose the lack of peer review at low-quality conferences that essentially scam researchers with publication and conference fees" [Bohannon, 2015].)

Now with generative AI platforms, creating machine-generated text and documents has become increasingly easy and accessible (Biswas, 2023). Additionally, the content is more realistic and, consequently, more difficult to detect in human peer review. For example, humans have been unable to detect scientific abstracts written by ChatGPT (Gao et al., 2023), which prompts broader questions about whether such technology applications can pass the Turing Test, aimed at testing AI's capacity for exhibiting intelligent behavior equivalent to or indistinguishable from that of a human.[7] On the one hand, such tools can generate legitimate research content. For instance, AI exclusively penned and published the book *Lithium-Ion Batteries* as Beta Author, and ChatGPT has "racked up at least four authorship credits on published papers and preprints."[8] At the same time, generative AI systems are far from perfect and are notorious for generating incorrect or unverifiable information, such as non-existent citations (Kendrick, 2023). They also can be (and are) used for nefarious purposes, such as to produce deep fakes, e.g., digital forgeries of scientific images (Farid, 2009).

---

[6] https://news.mit.edu/2015/how-three-mit-students-fooled-scientific-journals-0414

Here it is appropriate to distinguish between false negatives and false positives. Concerning peer review, the former refers to situations where "good" papers are rejected, and the latter "bad" papers are accepted.[9] Whether a human or a machine is involved in making a publication decision, each type of error has different implications for the legitimacy of peer review.

*Different types of errors pose different challenges for assessing the legitimacy and trustworthiness of AI in peer review.*

For instance, the first case could negatively impact scientific equity and equality — and efforts to broaden participation in science, especially for authors belonging to marginalized or underrepresented groups — which goes against democratic themes embedded in the scientific ethos. In the latter instance, where a fake study gets published, there can be profound adverse effects on society that span generations. For example, the infamous 1998 Wakefield study, which contained misinformation about the connection between vaccination and autism, is a case in point; it continues to fuel vaccine hesitancy and immunization-preventable disease outbreaks (Dubé et al., 2015).

AI authorship raises not only issues about the integrity of research and the peer review process, but also about accountability. Again, this matter has become especially pronounced with the release of ChatGPT and other generative AI platforms. In light of legitimacy concerns, publication guidelines more and more are including accountability as a criterion for authorship.[10] However, if an AI author is responsible for inaccuracies in a study or manuscript, should (and can) it be accountable? This could be a particularly thorny problem if the human author(s) do not understand how the technology arrived at the output or decision in the first place (de Silva, 2023), which is at odds with epistemic norms and other principles behind the ethos of science emphasizing openness and transparency. AI, as authors, also may be incongruent with communality, conflicting with the expectation that scientists should be accountable to their peers in literal terms. Additionally, it prompts higher-level philosophical questions about whether a machine constitutes an actor — a scientist — capable of thinking and reasoning like a human researcher.

Another threat to the legitimacy of peer review in the era of AI is the growing involvement of "big tech" in related activities. So-called "platform science" companies — i.e., the social media of the scientific enterprise (Mirowski, 2018) — rely heavily on AI algorithms for making recommendations and personalizing digital spaces. Such entities also are actively involved in designing AI applications for peer review and for scholarly communication more broadly. While for-profit organizations are benefitting from efforts to democratize scientific research by developing open and community-oriented platforms, their motivations may not align with social and scientific values and norms (Ortega, 2016), including those related to justice and fairness. In general, private sector interests will emphasize profit optimization and rates of return — and, in the case of social media, maximizing "eyeballs" for advertising purposes — over societal benefit (Schintler & McNeely, 2022), reflecting value tensions intrinsic to AI systems in the first place (Whittlestone et al., 2019). Similarly,

> *The involvement of private interest technology companies in scholarly communication raises the possibility that scientific processes ingrained in such platforms may belie epistemic interests, going against the norm of disinterestedness.*

---

[7] Based on work by Alan Turing in 1950. https://turing.org.uk/scrapbook/test.html
[8] https://www.nature.com/articles/d41586-023-00107-z#:~:text=The%20artificial%2Dintelligence%20(AI),on%20published%20papers%20and%20preprints
[9] https://transportist.org/2018/07/13/on-false-positives-and-false-negatives-and-peer-review
[10] E.g., those specified by the International Committee of Medical Journal Editors (ICMJE).

There are other ethical issues related to the online channels through which science circulates minute-to-minute, including mainstream social media. While such platforms have become commonplace for facilitating informal peer review, as discussed, AI "bots" and algorithms are pervasive in them. Bots are infamous for spreading mis- and dis-information, making "unwarranted beliefs" and "fake science"— i.e., science that is not "allayed by scientific evidence and authority" — to travel at the speed of light, a situation that is inconsonant with the scientific ethos, promoting "unorganized skepticism" rather than organized skepticism (Ezrahi, 2011). Coming back to the Wakefield study as an example, the internet (and social media) has permitted the fast diffusion of "anti-vaccination content," which is compounding problems related to vaccine hesitancy (Dubé et al., 2015).

An additional set of issues regarding the legitimacy of AI-supported peer review systems is that such applications tend to be opaque, with the code and architectural design known only to the developer. This situation is especially problematic when a company has developed the application, as its inner workings may not be disclosed for proprietary or other reasons (Schintler & McNeely, 2022).

> *AI's lack of transparency and interpretability is incongruent with epistemic norms and principles, as well as imperatives for science and related practices to be open and transparent.*

An example of a possible solution is Explainable Artificial Intelligence (XAI), which aims to make AI "more intelligible to humans by providing explanations" — e.g., "what it has done, what it is doing now, and what will happen next" — thus disclosing "salient information that it is acting on" (Gunning et al., 2019). However, XAI raises its own ethical issues that can militate against the values and norms embedded in the ethos of science (Schintler & McNeely, 2022). Even if rules and parameters embedded in an AI system are open, full transparency is not guaranteed; there is a difference between "seeing the whole code and understanding all of its potential effects" (Firth-Butterfield, 2017).

The increasing presence of technology companies in scholarly communication also raises issues concerning regulatory legitimacy. Such entities have enormous economic and political clout, so much so that they can self-regulate, which is inconsistent with democratic principles and ideals (Fukuyama et al., 2021). This also means that scientists are increasingly accountable to industry, thus compromising the autonomy of the scientific enterprise and conflicting with the norm of disinterestedness. While not a new situation in and of itself since science historically often has operated within a patronage system, autonomy is understood in relative terms; AI is used and subject to external regulation by the private sector, which could further undermine the sovereignty of the scientific enterprise.

### 3.2 Governance Roles and Implications

Another related matter concerns the globalizing production, commercialization, and application of AI technology, muddying the governance of AI-driven peer review by bringing multiple and potentially conflicting cultural norms, values, and laws into play (de Almeida et al., 2021), raising further questions about regulatory legitimacy. Furthermore, institutional change, such as legislative reform or cultural transformations, is slow compared to the speed at which technology advances. Current rules-in-use, normative understandings, and institutional imperatives are often out of sync with developing capabilities and applications of technologies, a problem referred to as "technological inversion" (Lessig, 20023). Polycentric governance, which invokes a flexible, adaptable, and localized strategy, rather than a rigid top-down regulatory regime, has been posited to address such complexities. This approach may be suitable for regulating AI peer review, given the polycentric nature of science in which many decision-making units have limited yet autonomous privileges and all operate under an overarching set of rules delineated in the scientific ethos. In any case, big questions arise here about "what" and "whose" values should be embedded in AI-driven peer review systems and how they should be governed in the first place, such that there is regulatory legitimacy.

AI is becoming more powerful and socially embedded (Gabriel & Ghazavi, 2021), which poses more threats to the legitimacy of peer review, especially considering that technology ultimately could be "uncontrollable and unpredictable by humans" (Krishna, 2020). Indeed, AI is "a growing resource of interactive, autonomous, and self-learning agency," which is "sufficiently informed, 'smart,'" and capable of performing morally "relevant actions independently of the humans who created them" (Floridi & Sanders, 2004), with implications for understanding and cognitive legitimacy. Generative AI platforms have heightened such concerns. Unlike prior AI systems, ChatGPT can learn and act "with little or no task-specific training" and minimal guidance from the user, arguably inching the technology closer to Artificial General Intelligence (AGI), where machines would be on par with humans "in every possible way" (Liebrenz et al., 2023). As such, using AI in peer review could further compromise the autonomy of the scientific enterprise, possibly making humans accountable not only to human peers but also to machines.

AI itself is in fact a regulatory system (Ågerfalk, 2020), encoded with governing norms, conventions, and dialogues, thus constraining and catalyzing institutional and social activities (Napoli, 2014). In this light, these systems embody values and principles in keeping with an organization's culture, reflecting its goals, desires, and expectations. *However, this perspective has become increasingly challenging as machines gain more agency and autonomy, potentially moving the needle even further away from legitimacy and the Mertonian ethos.*

## 4. Research and Policy Recommendations

This discussion speaks directly to problems of "value alignment" in AI, which focuses on ensuring that the technology's design, use, and application properly align with organizational and social values and expectations (Gabriel & Ghazavi, 2021), which also may conflict and invoke different aspects of legitimacy depending on perspective and purpose. Value alignment involves the engagement of moral and technical considerations based on intrinsic values and instrumental rationality, aiming to construct "virtuous technology" while also making it operational and practical (Friedman et al., 2013; Drori et al., 2003). In reference to peer review systems, several AI principles have been specified here as interpretive predicates generally grounded in values and norms related to respect for human autonomy, prevention of harm, and the promotion of fairness, explicability, and accountability, providing a framework for guiding AI design (Floridi & Cowls, 2022; Umbrello & Van de Poel, 2021). Such principles constitute a normative core — or ethos — for AI use in general (Fjeld et al., 2020), and more specifically here in application to peer review, also helping to steer the development of "laws, rules, technical standards, and best practices," with implications for legitimacy and trustworthiness in related use (Floridi & Cowls, 2022). At the same time, it is imperative to contextualize and gear AI principles to the applications at hand (Fjeld et al., 2020), i.e., to devise "culturally-sensitive" AI systems (Ferrara, 2023). However, the degree to which AI principles are congruent with the scientific ethos is unclear. Perhaps the ethos of science may need to evolve in light of or adapt to recent developments in AI. Thus, especially in consideration of the aforementioned norm-counternorm continuum, further research is needed to understand such complexities and dynamics, attending to effects within and across societal sectors and groups as discussed above. Ultimately, such efforts should strive to develop standards, guidelines, and other regulatory approaches that help ensure that the technical design of AI systems for peer review and related outcomes are legitimate in all possible ways.

  Gauging the perceptions and attitudes of the scientific community about using AI in peer review is crucial. Organizational legitimacy is the perceived appropriateness of an organization to a social system in terms of rules, values, norms, and definitions (Deephouse et al., 2017). However, while legitimacy is posited as an organizational or systems-level phenomenon, it is the legitimacy judgments of individuals collectively that "influence the norms, laws, and cognitive categories of social systems" (Martin & Waldman, 2022). Individual attitudes lead to legitimacy evaluations of organizational practices (Deephouse et al., 2017), and individual perceptions are context-specific, shaped by values acquired through experiences in different circumstances or domains of activity within a given community (Finch et al., 2015). In this sense, they are the micro-foundations of organizational legitimacy (Scott, 1995), i.e., the "micro-motor" of organizations (Powell & Colyvas, 2008). Accordingly, a better understanding is needed of individual normative judgments about algorithmically-driven peer review processes and outcomes, considering relevant contextual factors and dynamics. There is in fact a burgeoning area of research aimed at perceptions of the legitimacy of algorithmic decision-making, especially in relation to comparably situated human decision-making (Lima et al., 2021; Martin & Waldman, 2022; Starke & Lünich, 2020). However, perceptions about AI-driven peer review, particularly from within the scientific community, has been little examined (Tennant & Ross-Hellauer, 2020; Rowley & Sbaffi, 2018; Checco et al., 2020). Moreover, science encompasses unique structures, dynamics, and factors that drive perceptions about legitimacy. Different fields and disciplines have their own value systems, priorities, and cultures, including notions of what is legitimate or not. Thus, examining how such nuances affect perceptions of the legitimacy of AI use in peer review is of the essence.

  A significant priority for informing the formulation and evaluation of governance strategies regarding AI in peer review is the development of relevant metrics and evaluation benchmarks.

Such metrics should reflect, on the one hand, practical benefits of related applications and, on the other hand, ethical and social downsides and risks and broader implications for science and society. In doing so, it will be essential to determine a consistent and agreed-upon set of metrics for assessing key outcomes and processes, capturing different dimensions and dynamics of the issue. For example, to measure impacts related to justice and equity, aspects of both distributive fairness, emphasizing the equitable allocation of resources, and procedural fairness, or the "perceived fairness of the process that leads to the outcome" (Tomašev et al., 2020), should be considered. The construction and use of metrics are needed for evaluating the AI pipeline and all stages of the AI lifecycle — from the inception of systems to their final use and application — to ensure the proper consideration of values and principles associated with the scientific ethos. Moreover, metrics and strategies should enable the assessment of tradeoffs, e.g., among efficiency, accuracy, and fairness. Ultimately, such metrics must be sensitive to and account for contextual variations across research domains, geographies, and scientific communities, and also dynamically updated as the scientific ethos evolves, as it has historically vis-à-vis disruptive technological developments.

## 5. Conclusion

While we have discussed these issues relative to peer review, there is something much larger at the core of scholarly thought and engagement as a societal endeavor. The research here expands knowledge of scientific and also societal impacts involved with the growing use of AI in scientific peer review and scholarly communication more generally, and thus an understanding of the continuously changing relations of science, technology, and society (Krishna, 2020). This situation poses new challenges to existing norms, conventions, and expectations in scientific communities, such that this discussion has been positioned squarely in ongoing academic and scholarly dialogues and investigations about legitimate and trustworthy AI, especially in relation to AI-mediated peer review.

> *The point again is that the problems with AI are also problems with humans in the peer review process; this is an inherently sociological issue given that AI is embedded in social norms and practices. Research in this domain must keep that as a fundamental point of inquiry.*

Potentially affecting the nature of scientific work and its impacts on research productivity and culture, the deployment of generative AI systems and LLMs require particular policy attention in that they could "lead to more shallow work, blur concepts of authorship and ownership, and possibly create inequalities between speakers of high- and low-resource languages. However, LLMs and other forms of AI also could aid governance processes, such as in supporting peer review — an issue that requires further study (OECD 2023). With questions centered on whether humans or machines are more reliable and valid in their decisions, the emphasis is on the continuum of norms and counternorms in relation to the ethos of science. The analytical perspective here has encompassed ethical and social considerations emphasizing pragmatic, moral, regulatory, and cognitive aspects of legitimacy. Ultimately, the analysis provides a broader understanding of the legitimacy of scholarly communication processes — whether involving AI or humans — concerning trust in science, both inside and outside the ivory tower (NASEM, 2017).